%% file: arxiv_version.tex
\documentclass[sigconf]{acmart}

\setcopyright{none}
\renewcommand\footnotetextcopyrightpermission[1]{} 
\usepackage{booktabs}
\usepackage{tabularx}

\input{defines}

\begin{document}
\title{Computationally Inferred Genealogical Networks Uncover Long-Term Trends in Assortative Mating}

\author{Eric Malmi}
\affiliation{%
  \institution{Aalto University}
  \city{Espoo}
  \country{Finland}
}
\email{eric.malmi@aalto.fi}

\author{Aristides Gionis}
\affiliation{%
	\institution{Aalto University}
	\city{Espoo}
	\country{Finland}
}
\email{aristides.gionis@aalto.fi}

\author{Arno Solin}
\affiliation{%
  \institution{Aalto University}
  \city{Espoo}
  \country{Finland}
}
\email{arno.solin@aalto.fi}

\renewcommand{\shorttitle}{Computationally Inferred Genealogical Networks}

\begin{abstract}
Genealogical networks, also known as family trees or population pedigrees, are commonly studied by genealogists wanting to know about their ancestry, but they also provide a valuable resource for disciplines such as digital demography, genetics, and computational social science. These networks are typically constructed by hand through a very time-consuming process, which requires comparing large numbers of historical records manually. We develop computational methods for automatically inferring large-scale genealogical networks. A comparison with human-constructed networks attests to the accuracy of the proposed methods. To demonstrate the applicability of the inferred large-scale genealogical networks, we present a longitudinal analysis on the mating patterns observed in a network. This analysis shows a consistent tendency of people choosing a spouse with a similar socioeconomic status, a phenomenon known as \emph{assortative mating}. 
Interestingly, we do not observe this tendency to consistently decrease (nor increase) over our study period of 150 years.
\end{abstract}


\fancyhead{}


\maketitle

\section{Introduction}

\begin{figure}[t]
  \centering
  \includegraphics[width=\columnwidth]{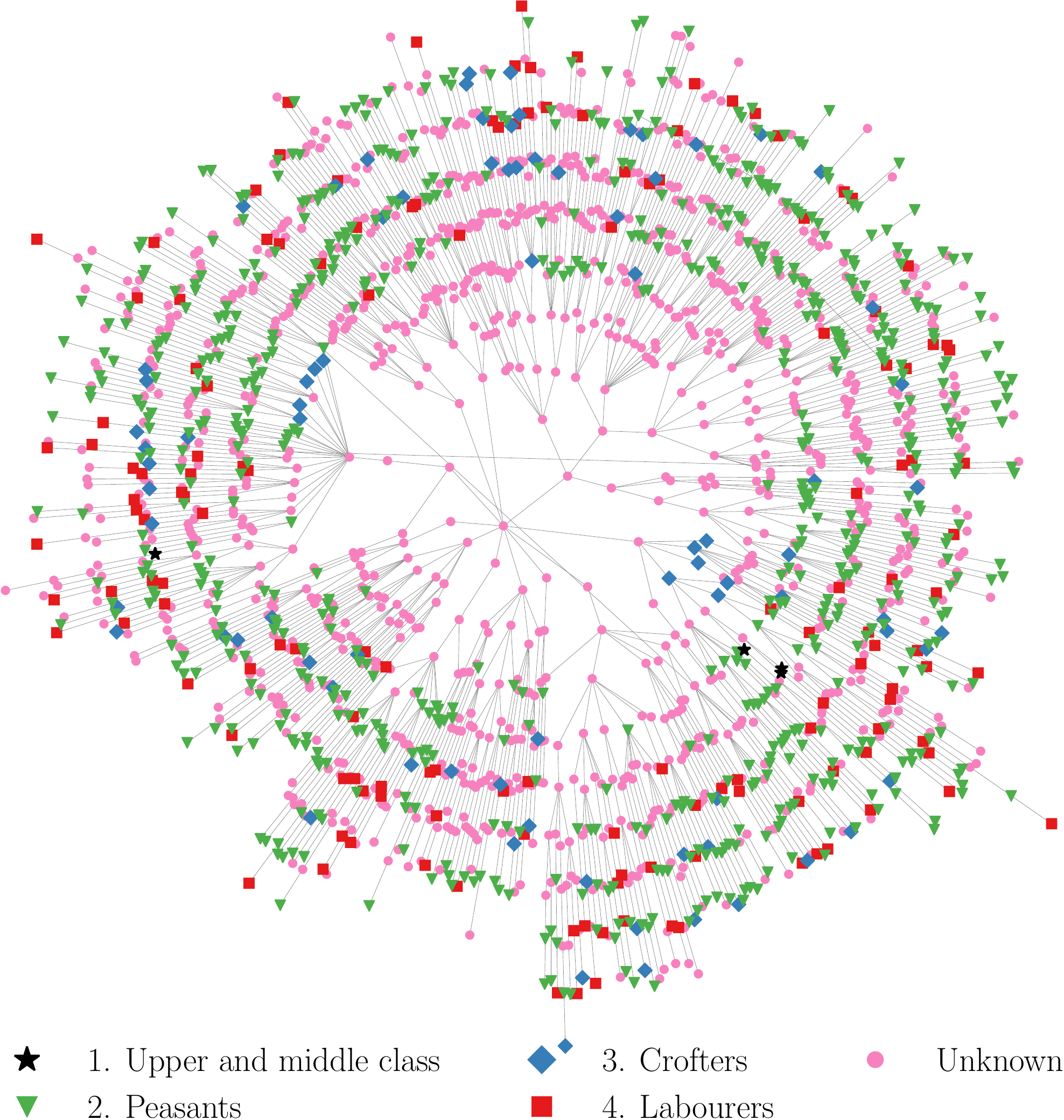}
  \caption{A subgraph of a genealogical network automatically inferred by 
  linking birth records. The subgraph spans 13 generations.\label{fig:tree}}
\end{figure}

\emph{Where do we come from? What are we? Where are we going?} These questions, posed by the famous painter Paul Gauguin, resonate with many people. An example of the allure of the first question can be attested in the popularity of genealogy, the study of family history. Genealogical research is typically conducted by studying a large number of historical vital records, such as birth and marriage records, and trying to link records referring to the same person. This process is nowadays facilitated by numerous popular online services, such as \textit{Ancestry.com}, \textit{MyHeritage}, and \textit{Geni}, and increasingly with genetic analysis. Nevertheless, constructing a genealogical network (also known as family tree or population pedigree) is still a very time consuming process, entailing lots of manual work.

In recent years, lots of efforts have been focused on \textit{indexing} 
historical vital records available in physical archives as well as in online 
repositories in the format of scanned images. Crowdsourcing projects organized, 
for example, by online genealogy services aim to do this kind of research by 
attracting people capable of interpreting old handwriting. Furthermore, there 
are also recent efforts at developing new optical character recognition (OCR) 
techniques to automate the process~\cite{giotis2017}. The availability of 
indexed records makes genealogical research amenable to new computer-supported 
and, to some extent, even fully-automatic approaches.

Our goal in this paper is to develop novel computational methods for inferring 
large-scale genealogical networks by linking vital records. We propose two 
supervised inference methods, \xgboost and \collective, which we train and test 
on Finnish data from the mid 17th to the late 19th century. More specifically, 
we apply these methods to link a large collection of indexed vital records 
from Finland, constructing a genealogical network whose largest component 
contains 2.6 million individuals. A small subgraph of this component is 
visualized in Figure~\ref{fig:tree}, where people are colored by their social 
class  
which is based on their father's occupation. The class information is used 
later when analyzing assortative mating. The construction of the full network 
takes 
only about one hour, which shows that it is vastly more scalable than the 
traditional manual approach which would probably require at least dozens of 
man-years for the same task. However, at its current stage, the accuracy of the 
automatic approach is not comparable to a careful human genealogist, but it can 
still support the work of the genealogist by providing the most probable 
parents for each individual, narrowing the search space.

From the methodological point of view, the main idea behind the proposed approaches is to cast the network-inference task into multiple binary classification tasks. Furthermore, our second approach, \collective, aims at capturing the observation that 
people tend to have children with the same partner, so if, for example, a 
father has multiple children, the mothers of the children should not be 
inferred independently. Incorporating this notion to our optimization problem, 
interestingly, leads to the well-known facility-location problem as well as an 
increase in the accuracy of the links.

In addition to family history, genealogical networks can be applied to several other domains, such as digital demography~\cite{weber2017}, genetics, human mobility, epidemiology, and computational social science. In this work, we demonstrate the applicability of the inferred network by analyzing \textit{assortative mating}, that is, a general tendency of people to choose a spouse with a similar socioeconomic background. Mating choices have been shown to be an important driver for income inequality~\cite{greenwood2014}.

From the perspective of computational social science~\cite{lazer2009}, genealogical networks offer particularly interesting analysis opportunities because of the long time window these networks cover. Compared to social-media data, which is typically used for computational social-science studies, genealogical networks contain less granular data about the people in the network, but they allow us to observe phenomena that occur over multiple generations. This aspect of genealogical networks enables us to quantify long-term trends in our society, as done for assortative mating in this work. Therefore, genealogical networks can provide answers not only to the first question asked in the beginning of this section but also to the third one: \textit{Where are we going?}

Our main contributions in this paper are summarized as follows:
\begin{itemize}
  \item We propose a principled probabilistic method, \xgboost, for inferring large-scale genealogical networks by linking vital records.
  \item We perform an experimental evaluation, which shows that 61.6\% of the links inferred by \xgboost are correct. The accuracy obtained by \xgboost surpasses the accuracy (56.9\%) of a recently proposed \naivebayes method~\cite{aai2017}. Furthermore, we show that the link probability estimates provided by \xgboost can be used to reliably control the precision--recall trade-off of the inferred network.
  \item We present a novel inference method, called \collective, which aims to improve disambiguation of linked entities, and does so by considering the genealogical-network inference task as a global optimization task, instead of inferring family relationships independently. \collective further improves the overall link accuracy to 65.1\%. However, contrasted to \xgboost, \collective does not provide link probabilities.
  \item Finally, we demonstrate the relevance and applicability of automatically inferred genealogical networks by performing an analysis on assortative mating. The analysis suggests that assortative mating existed in Finland between 1735 and 1885, but it did not consistently decrease or increase during this period.
\end{itemize}

\section{Data}

Our genealogical network inference method is based on linking vital records. The method is trained and evaluated using a human-constructed network. Next, we present these two data sources.

\subsection{Population Records}

The Swedish Church Law 1686 obliged the parishes in Sweden (and in Finland, 
which used to be part of Sweden) to keep records of births, marriages, and 
burials, across all classes of the society. The `HisKi' project, an effort 
started in the 1980s, aims to digitally index the hand-written Finnish parish 
registers. The digitized data contains about 5~million records of births and a 
total of 5~million records of deaths, marriages and migration. The HisKi 
dataset is publicly available at \url{http://hiski.genealogia.fi/hiski?en}, 
except for the last 100 years due to current legislation.

Each birth record typically contains 
the name, birth place, and birth date of the child in addition to the names and occupations of the parents. The goal of the genealogical network inference problem is to link the birth records to the birth records of the parents, creating a family tree with up to millions of individuals.

Currently, we only have access to data from Finland, but similar indexed datasets can be expected to become available for other countries through projects such as READ\footnote{\url{https://read.transkribus.eu/}} which develop optical character recognition (OCR) methods for historical hand-written documents.

\subsection{Ground Truth}
\label{sec:gt}

We have obtained a genealogical network consisting of 116\,640 individuals constructed by an individual genealogist over a long period of time. To use this network as a ground truth, we first match these individuals to the birth records in the HisKi dataset. An individual is considered matched if we find exactly one birth record with the same normalized first name and last name and the same birth date. Then, we find parent--child edges where both individuals are matched to a birth record, yielding 18\,731 ground-truth links.

Finally, the ground-truth links are split into a training set (70\%) and a test set (30\%). This is done by computing the connected components of the network, sorting the components by size in descending order, and assigning the nodes into two buckets in a round-robin fashion: First, assign the largest component to the training set. Second, assign the second largest component to the test set. Continue alternating between the buckets, however, skipping a bucket if its target size has been reached. Compared to directly splitting people into the two buckets, this approach ensures that there are no edges going across the two buckets that would thus be lost.
In the end, we get 5\,631 test links out of which 42\% are between mother and a child.

\section{Genealogical Network Inference}

Genealogical networks are typically constructed manually by linking vital records, such as birth, marriage, and death certificates. The main challenges in the linking process are posed by duplicate names, spelling variations and missing records.

\subsection{Problem Definition}
A \textit{genealogical network} is defined as a directed graph where the nodes correspond to people and the edges correspond to family relationships between them. We consider only two type of edges: \textit{father} edges, going from a father to a child, and \textit{mother} edges, going from a mother to a child. Each node can have at most one biological father and mother, but they are not necessarily known. Because of the temporal ordering of the nodes, this graph is a directed acyclic graph (DAG).

Each person in the graph is represented by the person's birth record. Given a set of birth records $V$, the objective of the genealogical network inference is to link each birth record $a \in V$ to the birth records of the person's mother $\Mom{a} \in V$ and father $\Dad{a} \in V$. In addition to the birth records, the inference method gets as input a set of mother candidates $\CandsMom \subseteq V$ and a set of father candidates $\CandsDad \subseteq V$ for each child. In each method studied in this paper, the candidate sets are defined as the people who were born between 10 and 70 years before the child and whose normalized first and last name match to the parent name mentioned in the child's record.
\footnote{The names are normalized with a tool developed by the authors of~\cite{aai2017}, which is available at: \url{https://github.com/ekQ/historical_name_normalizer}}

Since the true parents are ambiguous, the output should be a probability distribution over different parent candidates, including the case $\emptyset$ that a parent is not among the candidates.

\subsection{Naive Bayes Baseline}

This baseline method \cite{aai2017} first constructs an attribute similarity vector $\comps$ for each (child, candidate parent) pair. The vector consists of the following five features: a Jaro--Winkler name similarity for first names, last names and patronyms, and the age difference as well as the birth place distance between the child and the parent.

Assuming that the family links are independent, the probability of each mother 
candidate $m$ (and similarly of each father candidate) for person $a$ can be 
written as follows, using the Bayes' rule
\begin{equation}
 \pr{\Mom{a} = m \mid \comps} = \frac{\pr{\Mom{a}=m}\frac{\pr{\comp{a,m} \mid \Mom{a}=m}}{\pr{\comp{a,m} \mid \Mom{a} \neq m}}}{\sum_{m' \in \CandsMom_a \cup \emptyset} \pr{\Mom{a}=m'} \frac{\pr{\comp{a,m'} \mid \Mom{a}=m'}}{\pr{\comp{a,m'} \mid \Mom{a} \neq m'}}}, \label{eq:prob}
\end{equation}
where $\pr{\Mom{a}=m}$ is the prior probability of $m$ and $\pr{\comp{a,m} \mid 
\Mom{a}=m}$ denotes the likelihood of observing the attribute similarities 
$\comp{a,m}$.
The derivation of Equation~(\ref{eq:prob}) is given in the earlier work of~\citet{aai2017}, 
and we adopt it in this paper.

Since the links are assumed to be independent, the log-likelihood function over all links is given by the sum of the link log-probabilities. Let $x_{a,m} \in \{0, 1\}$ be a random variable denoting whether person $a$ is linked to parent $m$. This allows us to write the likelihood function~as
\begin{align}
	\max_{x} \quad & \bigg[ \sum_{a,m} \log\pr{\Mom{a}=m \mid \comps} x_{a,m} \nonumber \\
    & + \sum_{a',f} \log\pr{\Dad{a'}=f \mid \comps} x_{a',f} \bigg] \label{eq:objective} \\
    \textmd{such that}\quad & \sum_m x_{a,m} = 1, \quad \mbox{for all } a, \\
    & \sum_f x_{a,f} = 1, \quad \mbox{for all } a, \\
    & x_{a,m}, x_{a,f} \in \{0, 1\}, \quad \mbox{for all } a, m, f.
\end{align}
The maximum likelihood solution for this non-collective genealogical network 
inference problem can be obtained simply by optimizing the parent links 
independently.
The method assumes that the components of $\comp{a,m}$, corresponding to 
different attribute similarities, are also independent. 
Therefore, we call this method \naivebayes.

Next we present the two methods proposed in this paper, which improve over 
\naivebayes.

\subsection{Binary-Classification Approach}
The key observation behind this approach is that likelihood ratios can be 
approximated with probabilistic discriminative classifiers~\cite{cranmer2016}. 
This means that instead of estimating the component-wise probability 
distributions $\pr{\com{a,m}{i} \mid \Mom{a}=m}$, as done in \naivebayes, we 
can approximate the two likelihood ratios in (\ref{eq:prob}) by training a 
probabilistic binary classifier to separate attribute similarity vectors 
$\comp{a,m}$ corresponding to matching and non-matching (child, candidate 
parent) pairs. The most straightforward approximation is given by
\[
\frac{\pr{\comp{a,m} \mid \Mom{a}=m}}{\pr{\comp{a,m} \mid \Mom{a} \neq m}} \approx \frac{s\left(\comp{a,m}\right)}{1 - s\left(\comp{a,m}\right)},
\]
where $s\left(\comp{a,m}\right) \in [0,\, 1]$ is the output of a probabilistic binary classifier trained to separate matching $\comp{a,m}$ vectors from non-matching ones. In some cases, the probabilities predicted by the classifier can be distorted, which can be countered by calibrating the probabilities~\cite{niculescu2005predicting}.

We choose XGBoost~\cite{xgboost} as the classifier $s$ since it has been successfully employed for a record linkage task~\cite{tay2016,lian2016} as well as other classification tasks previously. In our case, the probabilities predicted by the XGBoost classifier are fairly accurate (calibration does not improve the Brier score \cite{brier1950verification} of the classifier) so calibration is not used. This approach is called \xgboost.

The key advantages that \xgboost offers over \naivebayes are: ($i$)~we do not 
have to make the independence assumption for attribute similarities, ($ii$)~we 
can use any existing classifier that provides probabilities as output, and 
($iii$)~it is very easy to add new features to the attribute similarity vector 
and retrain the model, whereas in \naivebayes we have to estimate two new 
likelihood distributions for each new feature, which requires manually 
selecting a suitable distribution based on the type of the new feature.

In total, we use 20 attribute similarity features $\comps$, which can be grouped into the following categories:
\begin{enumerate}
 \item \textbf{Candidate age} : The age of the candidate parent at the time of 
 the child's birth and the difference to the reported parent age if available 
 (for mothers, an approximated age is reported in 40\% of the birth records).
 \item \textbf{Geographical distance} : The distance between the birth places of the child and the candidate parent.
 \item \textbf{Names} : The similarity of the \textit{first} names, \textit{middle} names (if any), \textit{last} names, and \textit{patronyms} reported in the child's record for the parent and in the candidate's record for the child.
 \item \textbf{NaiveBayes} : The probability estimated by the \naivebayes method is used as a feature.
 \item \textbf{Gender} : A binary variable indicating whether we are matching a father or a mother.
 \item \textbf{Location} : The coordinates of the child's birth location.
 \item \textbf{Birth year} : The birth year of the child.
 \item \textbf{Candidate death} : A binary variable indicating whether there is a non-ambiguous death record, which indicates that the candidate had died before the birth of the child, and another variable indicating how long before the birth the death occurred (0 if it is not known to occur before the birth).
\end{enumerate}
Feature groups (6)--(8) as well as middle name and patronym similarities are not used by the \naivebayes model. Feature groups (6) and (7) are not useful on their own, since they are the same for each candidate parent of a child, but they might be informative together with other features. For instance, name similarity patterns might be time- or location-dependent.

To compute the features regarding the death year of the candidate (feature group (8)), we first match the death records to birth records by finding the record pairs that ($i$)~have the same normalized first name, last name, and patronym, ($ii$) for which the reported age at death matches to the birth year $\pm 1$ year, and ($iii$)~the birth and the death location are at most 60 kilometers apart. Then we link a death record to a birth record only if the death record has only one feasible match. This approach allows us to infer the death time for 12\% of the birth records.\footnote{Based on the ground-truth data (Section~\ref{sec:gt}), 96.4\% of the obtained death record matches are correct, but we are able to match only 18\% of the 3.3 million death records. Ideally, the death record matching should be done in a probabilistic way, jointly with the birth record linking.}

\subsection{Collective Approach}

\begin{figure}[t]
  \centering
  \includegraphics[width=\columnwidth]{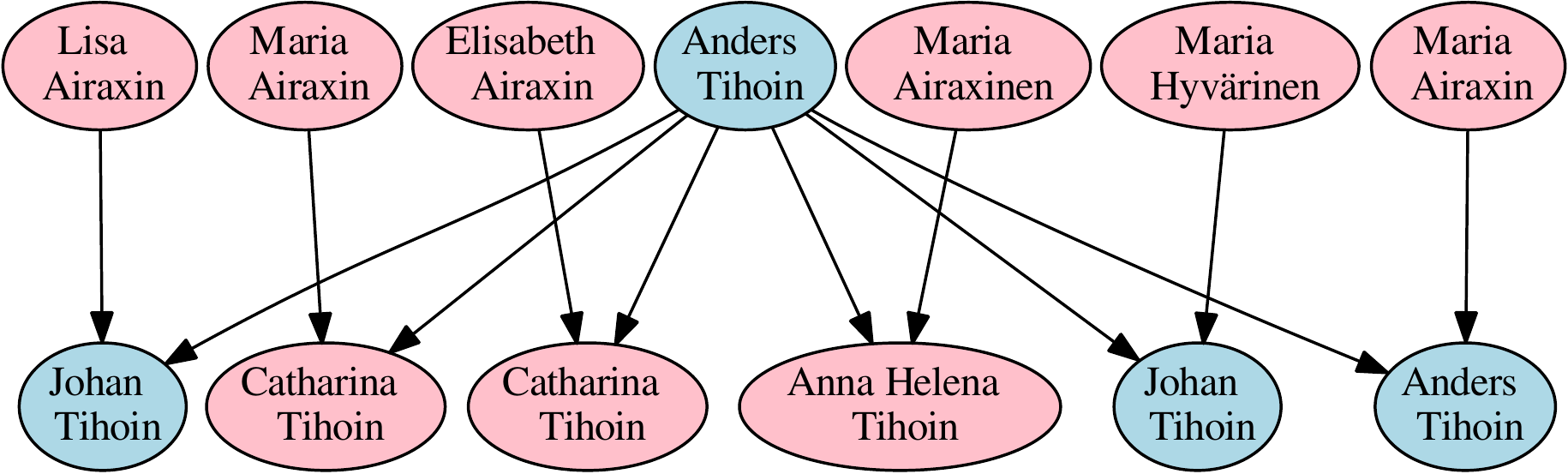}
  \caption{A sample subnetwork inferred by \naivebayes with colors corresponding to gender. If each child is matched to the most probable father--mother pair independently, the number of spouses per person can be unrealistically high.\label{fig:failure}}
\end{figure}

Both \naivebayes and \xgboost assume that the family links are independent, 
which leads to some unlikely outcomes. In particular, the number of spouses per 
person becomes unrealistically high which is illustrated in 
Figure~\ref{fig:failure} which shows a subgraph inferred by \naivebayes. The 
algorithm has inferred a person called \textit{Anders Tihoin} to have fathered 
six children---each with a different mother. While this is possible, it is very 
likely that at least the mothers of \textit{Catharina} (the left one), 
\textit{Anna Helena} and \textit{Anders} are actually the same person, since 
the mother's name for these children is almost the same \textit{(Maria 
Airaxin(en))}.

To address this problem, we propose to minimize the number mother--father pairs in addition to maximizing the probability of the inferred links. Let $y_{m,f} \in \{0, 1\}$ indicate whether any child has been assigned to mother--father pair $(m, f)$. Now we can write the collective genealogical network inference problem as
\begin{align}
	\max_{x,y} \quad & \bigg[- \lambda \sum_{m,f} y_{m,f} + \sum_{a,m} \log\pr{\Mom{a}=m \mid \comps} x_{a,m} \nonumber \\
    & + \sum_{a',f} \log\pr{\Dad{a'}=f \mid \comps} x_{a',f} \bigg] \label{eq:collective} \\
    \textmd{such that}\quad & \sum_a x_{a,m} x_{a,f} \leq y_{m,f}, \quad \mbox{for all } m, f, \\
    & \sum_m x_{a,m} = 1, \quad \sum_f x_{a,f} = 1, \quad \mbox{for all } a, \\
    & x_{a,m}, x_{a,f}, y_{m,f} \in \{0, 1\}, \quad \mbox{for all } a, m, f,
\end{align}
where $\lambda \geq 0$ controls the penalty induced by each extra parent pair (or discount for merging two parent pairs into one).

This optimization problem is an instance of the \textit{uncapacitated facility-location problem}, where parent pairs correspond to facilities, child nodes to demand sites, and the parameter $\lambda$ to the facility opening cost. The uncapacitated facility-location problem is \np-hard for general graphs so we adopt a greedy approach presented in Algorithm~\ref{alg:greedy}. Assuming that each person has $c$ candidate mothers and fathers, the time complexity of this algorithm is $\mathcal{O}\left(c^2|V| + |V| \log |V|\right)$, which shows that it scales well in the number of people to be linked. The input probabilities $p$ are computed with \xgboost. Note that the facility assignment costs, that are based on the probabilities, are not necessarily a metric, so the approximation guarantees for methods, such as \cite{jain2001}, do not necessarily hold in our problem setting.

\begin{algorithm}[t]
  \begin{algorithmic}[1]
      \Statex {\bf Input:} Child nodes $V$, parent candidates $\CandsMom, \CandsDad$ and their probabilities $p$, parameter $\lambda$.
      \Statex {\bf Output:} Inferred parent--child links $R$.
      \State $R = \emptyset$
      \State $Q = \emptyset$ \Comment{Set of used mother--father pairs.}
      \State Sort children $a \in V$ by $\max_{m,f} \log\pr{\Mom{a}=m \mid \comps} + \log\pr{\Dad{a}=f \mid \comps}$ in a descending order.
      \For{$a \in V$}
        \State $p_\textmd{max} = -1$  \Comment{Maximum probability.}
        \State $m_\textmd{max} = -1$  \Comment{Best mother.}
        \State $f_\textmd{max} = -1$  \Comment{Best father.}
        \For{$(m,f) \in \CandsMom_a \times \CandsDad_a$}
          \State $p_{m,f} = \log\pr{\Mom{a}=m \mid \comps} + \log\pr{\Dad{a}=f \mid \comps}$
          \If{$(m,f) \notin Q$}
            \State $p_{m,f} = p_{m,f} - \lambda$
          \EndIf
          \If{$p_{m,f} > p_\textmd{max}$}
            \State $p_\textmd{max} = p_{m,f}$
            \State $m_\textmd{max} = m$
            \State $f_\textmd{max} = f$
          \EndIf
        \EndFor
        \State $R = R \cup (m_\textmd{max},a) \cup (f_\textmd{max}, a)$
        \State $Q = Q \cup (m_\textmd{max}, f_\textmd{max})$
      \EndFor
      \State {\bf return} $\mathbf{R}$
  \end{algorithmic}
  \caption{\label{alg:greedy} A greedy method for collective genealogical network inference.}
\end{algorithm}

This method is called \collective. It outputs a genealogical network, where 
some of the links inferred by \xgboost have been rewired in order to reduce the 
number of spouses. A limitation of \collective is that it does not recompute 
the link probabilities (the marginal distributions of the random variables $x$, 
indicating the parents). One approach for computing the marginal distributions 
would be to adopt a Markov chain Monte Carlo (MCMC) method, which samples 
genealogical networks by proposing swaps to the parent assignments. However, to 
obtain a meaningful level of precision for the link probability estimates we 
need to sample a very large number genealogical networks,
which renders such an MCMC approach non-scalable. Thus, in this work, we are 
not experimenting with link-probability estimation for \collective.

\section{Experimental Evaluation}
\begin{figure*}[t!]
    \begin{subfigure}[t]{0.33\textwidth}
        \centering
        \includegraphics[width=\textwidth]{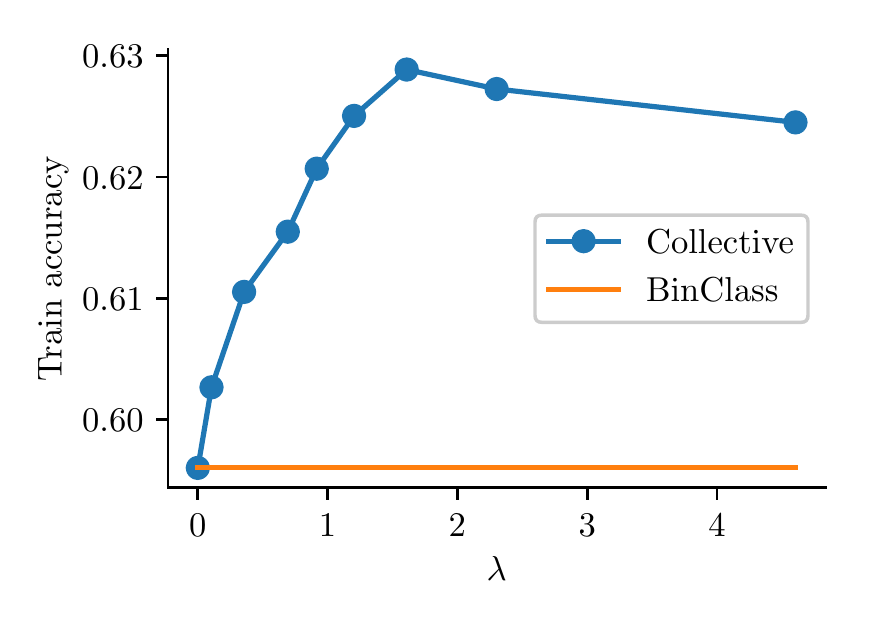}
	    \caption{Optimizing parameter $\lambda$ which controls the penalty induced by each extra parent pair in the \collective linking method. \label{fig:fl}}
    \end{subfigure}\hfill
    \begin{subfigure}[t]{0.28\textwidth}
        \centering
        \includegraphics[width=\textwidth]{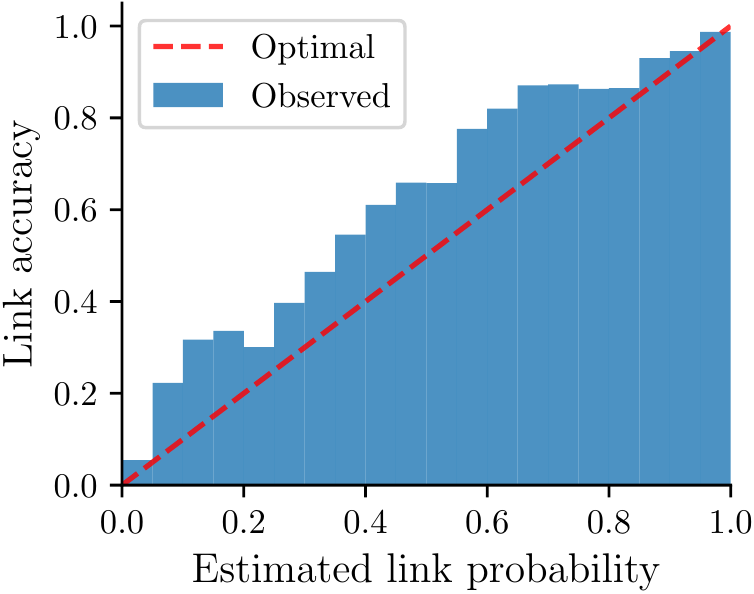}
        \caption{The link-probability estimates by \xgboost correlate strongly with the accuracy of the links binned by their probability. \label{fig:probs}}
    \end{subfigure}\hfill
    \begin{subfigure}[t]{0.33\textwidth}
        \centering
        \includegraphics[width=\textwidth]{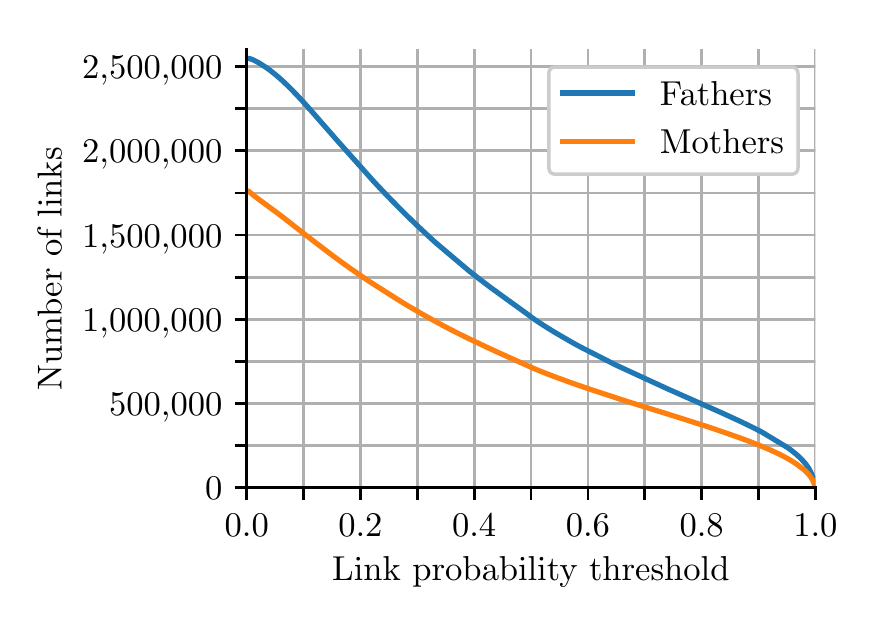}
        \caption{The number of inferred family links with the estimated link probability above a given threshold. \label{fig:recall}}
    \end{subfigure}
    \caption{Experimental results on genealogical network inference.}
\end{figure*}

Next, we compare the proposed methods, \xgboost and \collective, to two baseline methods, \naivebayes \cite{aai2017} and \random, the latter of which randomly assigns the parents among the set of candidates. The methods are evaluated by computing their \textit{link accuracy} which is the fraction of ground-truth child--parent links correctly inferred by the method. To use \collective, we first need to optimize $\lambda$---the penalty induced by each extra parent pair. The optimization is done using the training ground-truth data and the results are shown in Figure~\ref{fig:fl}. The training accuracy is maximized when $\lambda = 1.6$.

The results for all methods are presented in Table~\ref{tab:results}. \xgboost clearly outperforms the two baseline methods, and \collective further improves the accuracy of \xgboost from 61.6\% to 65.1\%.

\begin{table}[t]
\caption{Accuracy of the links inferred with different methods.}\label{tab:results}
\centering
\resizebox{\columnwidth}{!}{%
\begin{tabular}{lllll}
\toprule
\textbf{Method} & \random & \naivebayes & \xgboost & \collective \\
\textbf{Accuracy} & 12.5\% & 56.9\% & 61.6\% & \textbf{65.1\%} \\
\bottomrule
\end{tabular}%
}
\end{table}

To evaluate the accuracy of the link probabilities estimated by \xgboost, we bin the probabilities and compute accuracy within each bin. The results presented in Figure~\ref{fig:probs} show that the estimated link probabilities are somewhat pessimistic, but overall, they are well in line with the link accuracies. Therefore, we can use the link probabilities to filter out links below a desired certainty level when analyzing the inferred network, as done in the next section. This filtering improves the precision of the inferred network but naturally also decreases the recall as illustrated in Figure~\ref{fig:recall}. In total, \xgboost finds 1.8 million child--mother links and 2.5 million child--father links.
If we require a minimum link probability of 90\%, as done in the next section, 
we are still left with 253\,814 child--mother links and 341\,010 child--father 
links.

\begin{figure*}[t]
  \centering
  \includegraphics[width=\textwidth]{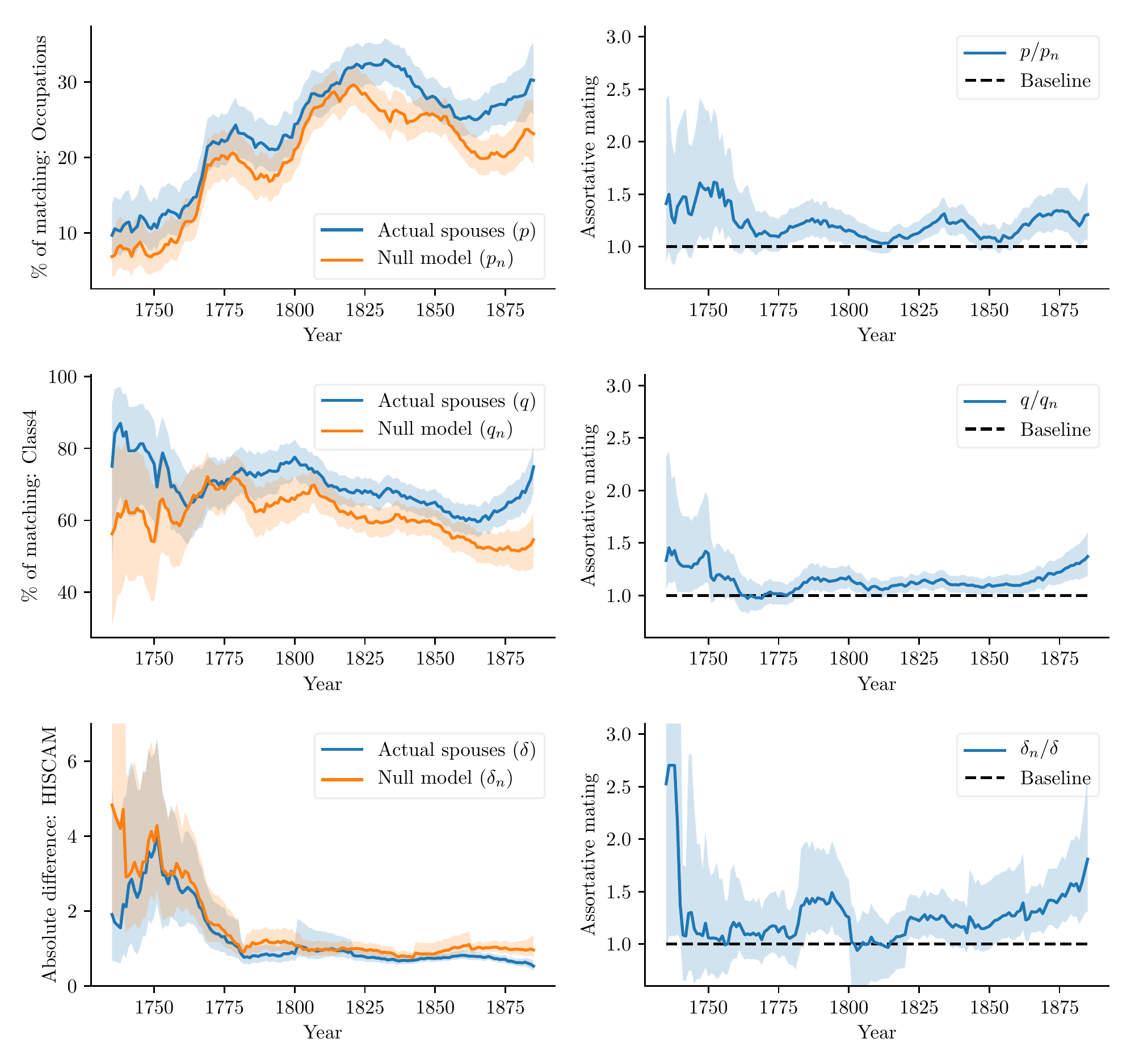}
  \caption{Assortative mating is detected in the inferred genealogical networks for Finland (1735--1885), but the phenomenon is not monotonously decreasing or increasing. Shaded areas correspond to the 95\% bootstrap confidence intervals. \label{fig:strat}}
\end{figure*}

\section{Case Study: Assortative Mating}
\label{sec:case}

\textit{Assortative mating}, also known as \textit{social homogamy}, refers to the phenomenon that people tend to marry spouses with a similar socioeconomic status and it is one instance of \textit{social stratification}. A recent study shows that assortative mating contributes to income inequality and it has been on the rise between 1960 and 2005~\cite{greenwood2014}. In this section, we leverage the inferred genealogical network to address two questions:
\begin{itemize}
 \item[($i$)] Can we detect assortative mating in historical Finland?
 \item[($ii$)] How has the intensity of assortative mating evolved in historical Finland?
\end{itemize}

\subsection{Estimating Social Status}

To detect assortative mating, we compare the socioeconomic status of the spouses inferred by our linking method. We use \textit{occupation} as a proxy for status, and rather than comparing the occupations of the spouses directly, we compare the occupations of the spouses' fathers. The father occupations are more comparable since occupations were strongly gendered in the 18th and 19th centuries, which are the focus of this study. Furthermore, there is a separate field for father occupation in the birth records so we do not need to separately infer the fathers but only the spouses.

The numerous variants and alternative abbreviations of the same occupation pose a challenge when comparing occupations. For instance, \textit{farmer}, which is the most common occupation in the dataset, goes by the following (non-comprehensive) list of titles: \textit{bonden, bd., b., bd:, b:, b:n, bdn, b:den, talollinen, talonpoika, tal., talp., tl., tln., talonp}.
To address this challenge, we normalize the occupation titles by first removing special characters and then comparing the title to lists of abbreviations available online~\cite{historismi1,historismi2}. We have also manually normalized the most common abbreviations not found in the online lists to increase the coverage of the normalization. In total, we are able to normalize 77.5\% of all non-empty father occupation titles found in the birth records.

In addition to comparing the normalized occupations directly, we also map them to the historical international classification of occupations (HISCO) \cite{hisco}. Then we divide the HISCO classes into four main classes: (1)~upper and middle class, (2)~peasants (who own land), (3)~crofters (who rent land), and (4)~labourers (who live at another person's house). The class to which the father of a person belongs to is called \classF. We also map the HISCO classes into an occupational stratification scale \hiscam~\cite{lambert2013}.\footnote{More specifically, we use the ``U2: Male only, 1800--1938'' scale. For more information on the different scales, see: \url{http://www.camsis.stir.ac.uk/hiscam/}} \hiscam is a real-valued number between 0 and 100 which measures the social interaction distance of people based on their occupations. The \classF and \hiscam codes are obtained for 75.5\% of the people.

\subsection{Measuring Assortative Mating}

A high percentage of matching spouse father occupations (\pM) is a signal of assortative mating but this percentage might also be affected by external factors such as the number of distinct occupations people had at a given time in a given city or the availability of data from different cities at different time periods. To control for these external factors, we introduce a null model, which shuffles the spouses within a city and a time window of 20 years and then computes the percentage of matching occupations (\pN). Then we measure assortative mating as the ratio of the two percentages ($\pM / \pN$). This ratio measures how much more likely people marry someone with a similar social status compared to a null model where the marriages are randomized. Thus a ratio larger than one is a sign of assortative mating.

The higher-level occupation categories \classF can also be used to compute 
match percentages ($\qM$ and  $\qN$) and their ratio $\qM / \qN$. With the 
\hiscam scores, we compute the mean absolute difference of the scores for 
the spouses (\dM) and for the null model (\dN), instead of the percentages. The 
assortative mating measure, in this case, is defined as $\dN / \dM$, so that 
again, a ratio larger than one indicates assortative mating.

\subsection{Data}

As the set of spouses to be analyzed, we use all pairs of people who have been 
inferred to be the mother and the father of the same child. Both parent links 
are required to have at least a 90\% probability (this threshold is varied in 
Appendix~\ref{sec:appendix}). 
We limit the analysis to the period with the most records from 1735 to 1885. Requiring both parents to have a sufficiently high link probability limits the number of spouses to 14\,542 pairs. Out of these, the father occupation is known for both spouses in 6\,402 pairs and \classF and \hiscam in 3\,128 pairs. Although these filtering steps significantly reduce the number of spouses, we are still left with a sufficiently large dataset to perform a longitudinal analysis on assortative mating.

\subsection{Results}

The percentages of matching spouse father occupations over time and the corresponding assortative mating curve are shown on the top row of Figure~\ref{fig:strat}. In order to highlight long-term trends, the curves show moving averages where a data point at year $y$ uses the inferred spouses from years $[y-10, y+10]$ (the effect of varying the delta value of 10 years is studied in Appendix~\ref{sec:appendix}). The figure also shows the 95\% bootstrap confidence intervals.

The middle and the bottom rows show the corresponding curves for the occupations grouped into four main classes (\classF) and for the numerical \hiscam scores of the occupations, respectively.

All three measures of assortative mating suggest that assortative mating did occur in Finland between years 1735 and 1885 since the assortative mating curves are fairly consistently above the baseline ratio of 1. The intensity of the phenomenon varies mainly between 1 and 1.5 but interestingly, there is no monotonically decreasing or increasing trend.

Finally, we observe that the spouse similarity curves \pM, \qM and \dM have very different shapes, whereas the three assortative mating curves are clearly correlated (the correlation coefficients are 0.69, 0.23, and 0.62 for pairs ($\pM / \pN$ vs.\ $\qM / \qN$), ($\pM / \pN$ vs.\ $\dN / \dM$), and ($\qM / \qN$ vs.\ $\dN / \dM$), respectively). This suggests that the proposed measures of assortative mating, which account for a null model, measure the phenomenon robustly.

\section{Related Work}

Genealogical network inference, also known as \textit{population reconstruction}~\cite{bloothooft2015}, is one application of \textit{record linkage}, which has been an active research area for many decades and was mentioned already in 1946 by Halbert L.\ Dunn~\cite{dunn1946}---interestingly, in the context of linking birth, marriage, and other vital records. Two decades later, Fellegi and Sunter published a widely cited paper on probabilistic record linkage~\cite{fellegi1969}. This approach considers a vector of attribute similarities between two records to be matched and then computes the optimal decision rule (\textit{link} vs.\ \textit{possible link} vs.\ \textit{non-link}), assuming independent attributes. The \xgboost method, proposed in this work, adopts a similar approach, however, employing a supervised classifier, which has the advantage of capturing dependencies between variables.

Record linkage has been mostly studied for other applications, but recently, with the rise in the number of indexed genealogical datasets, several studies have applied record linkage techniques for genealogical data~\cite{efremova2015,christen2015b,christen2016,kouki2016,kouki2017,malmi2017lagrangian,malmi2017,ranjbar2015hider}. The most closely related to our work are the papers by Efremova et~al.~\cite{efremova2015}, Christen~\cite{christen2016}, and Kouki et~al.~\cite{kouki2017}.

Efremova et~al.~\cite{efremova2015} consider the problem of linking records 
from multiple genealogical datasets. They cast the linking problem into 
supervised binary classification tasks, similar to this work, and find name 
popularity, geographical distance, and co-reference information to be important 
features. In our approach, we can avoid having to explicitly model name 
popularity, since the probability of a candidate parent is normalized over the 
set of all candidate parents, 
and for popular names this set will be large, thus downweighting the probability of the candidate.

Christen~\cite{christen2016} and Kouki et~al.~\cite{kouki2017} propose 
collective methods for linking vital records. The former method is evaluated on 
historical Scottish data and it is concluded that due many intrinsically 
difficult linking cases, the results are inferior to a linkage constructed by a 
domain expert. The latter method is evaluated on a more recent dataset 
collected by the National Institutes of Health and it yields a high linking 
F-measure of 0.946. This method is based on probabilistic soft logic 
(PSL), which makes it possible to add new relational rules to the 
model without having to 
update the inference method. In comparison to these works, we evaluate our 
methods on a dataset that is two orders of magnitude larger, and we show that 
the proposed \xgboost method not only provides accurate matches but also 
reliably quantifies the certainty of the matches---an important feature of a 
practical entity-resolution system.

Assortative mating,\footnote{Assortative mating can also refer to 
\textit{genetic} assortative mating, but here we use it exclusively to refer to 
sociological assortative mating.} also known as \textit{social homogamy}, has 
been studied widely in the sociology literature. The phenomenon has a 
significant societal impact since it has been shown to be connected to income 
inequality~\cite{greenwood2014}.
Greenwood et.~al.~\cite{greenwood2014} also find that assortative mating has 
been on the rise 1960--2005. Bull~\cite{bull2005} studies assortative mating in 
Norway, 1750--1900, focusing on farmers and farm workers. They find that 
assortative mating was declining in the mid-1700s but after that it stayed 
fairly constant. This finding is largely in line with our results for Finland 
from the same time period. To control for the variations in the sizes of the 
groups, Bull computes odds ratios~\cite{bull2005,kalmijn1998}, whereas we use a 
null model, which randomizes spouses. 

\section{Discussion and Conclusions}

We presented a principled probabilistic machine-learning approach, \xgboost, for inferring genealogical networks. This approach was applied to a dataset of 5.0~million birth records and 3.3 million death records based on which it inferred a network, containing 13 generations and a connected component of 2.6~million individuals, taking about one hour when parallelized over 50 machines. A comparison against a large human-compiled network yielded a link accuracy of 61.6\%, outperforming a naive Bayes baseline method. We showed that the accuracy can be further improved to 65.1\% by a collective approach called \collective.

A key feature of \xgboost is that it outputs probabilities for the inferred links. This allowed us to separate links that have a sufficiently high probability and to perform an analysis on the mating patterns observed in the network. The main findings of the analysis are that: ($i$)~assortative mating---the tendency to select a spouse with a similar socioeconomic status---did occur in Finland, 1735--1885, and ($ii$)~assortative mating did not monotonically decrease nor increase during this time period.

\spara{Limitations and future work.} There is always uncertainty involved when analyzing records that are several centuries old and even an experienced domain expert can make incorrect linking decisions. Since our method relies on human-generated training data, any mistakes or biases in the training dataset affect the resulting model and potentially the downstream analyses. Although the automatic approach enables data generation for various analyses at an unprecedented scale, one has to be careful when drawing conclusions. For instance, an interesting phenomenon that could be studied with the inferred networks would be long-term human mobility patterns. However, since the model uses features that are based on location, any potential biases learned by the model would directly affect the analysis of the resulting mobility patterns.

Many exciting future directions are left unexplored. 
First, it would be useful to extend the problem formulation so that it tries to jointly link all available record types, as now we only link birth records, using some features based on death records linked to the births separately.
Second, it would be interesting to incorporate additional collective terms to 
the optimization problem, capturing patterns such as namesaking (a practice to 
name a child after an ancestor) or constraints on the family relations between 
individuals, which could be obtained from DNA tests. Third, there are many 
options for extending the assortative mating analysis presented in this work. 
We could study the differences in the strength of this phenomenon between 
social classes and between, for example, the first and the latter children. We 
could also extend the analysis to a related phenomenon of \textit{social 
mobility}, which considers the status differences between parents and children 
instead of spouses. Fourth, with the advent of improved handwritten text 
recognition methods, similar historical birth record datasets can be expected 
to become available for many other countries. This will enable studying the 
generalizability of the proposed methods and the assortative mating analysis 
beyond Finland.
Finally, the presented inference and analysis methods could be applied to other types of genealogical networks such as the genealogical networks of Web content \cite{baeza2008genealogical}.

To facilitate future research on these and other related research problems, the 
data and the code used in this paper have been made available at: 
\url{https://github.com/ekQ/genealogy}.

\begin{acks}
We would like to thank the Genealogical Society of Finland for providing the 
population-record data, Pekka Valta for the ground-truth data, and Antti 
H{\"a}kkinen for the HISCO and \classF mappings. We are also grateful for 
useful discussions with Przemyslaw Grabowicz, Sami Liedes, Jari Saram{\"a}ki, 
Ingmar Weber, and Emilio Zagheni.
We also acknowledge the computational resources provided by the Aalto 
Science-IT project. Eric Malmi and Aristides Gionis were supported by the EC 
H2020 RIA project ``SoBigData'' (654024).
\end{acks}

\appendix
\section{Assortative Mating Sensitivity Analysis}
\label{sec:appendix}

\begin{figure*}[t]
  \centering
  \includegraphics[width=\textwidth]{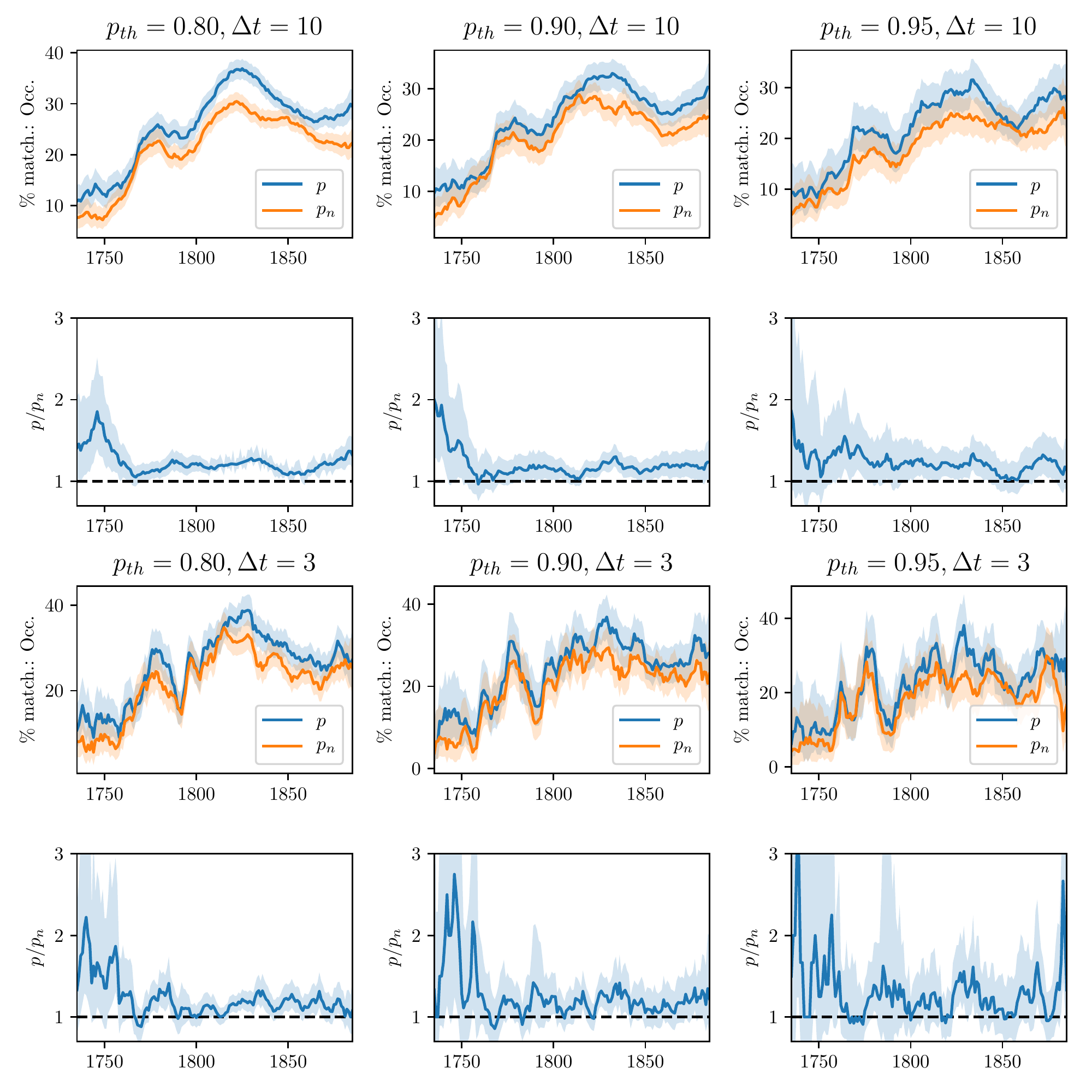}
  \caption{A sensitivity analysis showing the effect of varying the minimum link probability threshold ($p_{th}$) and the moving average delta ($\Delta t$) on the assortative mating curves from Figure~\ref{fig:strat} (top row). \label{fig:sa}}
\end{figure*}

In this section, we study the effect of varying the minimum link probability threshold ($p_{th}$) of 90\% and the moving average delta ($\Delta t$) of 10 years used in the analysis of assortative mating in Section~\ref{sec:case}. In the interest of space, we consider only the first assortative mating measure $\pM/pN$, which is based on the normalized occupations. The results are shown in Figure~\ref{fig:sa}.

First, we note that the larger the threshold $p_{th}$ or the smaller the smoothing parameter $\Delta t$, the less data points we have per year and thus larger the confidence intervals. Second, varying the threshold does not seem to affect the $\pM/pN$ curves significantly. Third, decreasing the smoothing parameter to $\Delta t = 3$ makes the curves considerably less smooth, which suggests that a higher smoothing parameter might highlight the long-term trends more effectively.

We conclude that these results support the findings that assortative 
mating did occur in Finland, 1735--1885, but it did not monotonically decrease 
nor increase during this time period.

\balance
\bibliographystyle{ACM-Reference-Format}
\bibliography{sample-bibliography}

\end{document}

%% file: defines.tex
\usepackage{suffix}
\usepackage{color}
\usepackage{dsfont}
\usepackage{url}
\usepackage{xspace}
\usepackage{amsfonts}
\usepackage{amsmath}
\usepackage{algorithm}
\usepackage[shortlabels]{enumitem}
\usepackage[noend]{algpseudocode}
\usepackage{subcaption}
\usepackage{balance}

\newcommand{\Mom}[1]{\ensuremath{M_{#1}}\xspace}
\newcommand{\Dad}[1]{\ensuremath{F_{#1}}\xspace}
\newcommand{\CandsMom}{\ensuremath{C^\textmd{m}}\xspace}
\newcommand{\CandsDad}{\ensuremath{C^\textmd{f}}\xspace}
\newcommand{\pr}[1]{p\left(#1\right)}
\newcommand{\naivebayes}{{\sc NaiveBayes}\xspace}
\newcommand{\xgboost}{{\sc BinClass}\xspace}
\newcommand{\collective}{{\sc Collective}\xspace}
\newcommand{\random}{{\sc RandomCand}\xspace}

\newcommand{\comps}{\ensuremath{\boldsymbol{\gamma}}\xspace}
\newcommand{\comp}[1]{\ensuremath{\comps^{#1}}\xspace}
\newcommand{\com}[2]{\ensuremath{\gamma^{#1}_{#2}}\xspace}

\newcommand{\dM}{\ensuremath{\delta}\xspace}
\newcommand{\dN}{\ensuremath{\delta_n}\xspace}
\newcommand{\pM}{\ensuremath{p}\xspace}
\newcommand{\pN}{\ensuremath{p_n}\xspace}
\newcommand{\qM}{\ensuremath{q}\xspace}
\newcommand{\qN}{\ensuremath{q_n}\xspace}

\newcommand{\classF}{{\sc Class4}\xspace}
\newcommand{\hiscam}{{\sc HISCAM}\xspace}

\newcommand{\spara}[1]{\smallskip\noindent{\bf{#1}}}

\newcommand{\fpr}[1]{\mathopen{}\left(#1\right)}

\newcommand{\np}{\textbf{NP}}

\DeclareRobustCommand{\dispfunc}[2]{%
  \ensuremath{%
  \ifthenelse{\equal{#2}{}}%
    {{#1}}%
    {{#1}\fpr{#2}}}}

\newcommand{\dd}{\ensuremath{K}}

\newcommand{\dist}[1][(p, q)]{\def\ArgI{{#1}}\distRelay}
\newcommand{\distRelay}[2][]{\dispfunc{\dd^{\ArgI}_{#1}}{#2}}
\WithSuffix\newcommand\dist*[2][]{\dist[][#1]{#2}}